 \documentstyle[preprint,aps]{revtex}
 
\begin{document}

 \draft

 \title{Real--space renormalization group for the\\
 random--field Ising model}
 \author{M. E. J. Newman, B. W. Roberts, G. T. Barkema, and J. P. Sethna}
 \address{Laboratory of Atomic and Solid--State Physics,\\
 Cornell University,\\
 Ithaca.  NY 14853--2501.}
 \maketitle

\begin{abstract}
We present real--space renormalization group (RG)
calculations of the critical properties of the random--field Ising model on
a cubic lattice in three dimensions.  We calculate the RG flows in a
two--parameter truncation of the Hamiltonian space.  As predicted, the
transition at finite randomness is controlled by a zero temperature,
disordered critical fixed point, and we exhibit the universal crossover
trajectory from the pure Ising critical point.  We extract scaling fields
and critical exponents, and study the distribution of barrier heights
between states as a function of length scale.
\end{abstract}

\vspace{1cm}

 \narrowtext


The critical properties of the random--field Ising
model~\cite{belanger,nattermann} have been the subject of intense
controversy.  One of the simplest of disordered systems, the model is
governed by the Hamiltonian
 \begin{equation}
 {\cal H} \equiv -J\sum_{\langle ij \rangle} s_i s_j - \sum_i (H+h_i) s_i,
 \end{equation}
where $s_i=\pm1$ is an Ising spin on a cubic lattice, $H$ is a homogeneous
external field and the variables $h_i$ are independent
Gaussianly-distributed random fields of mean zero and variance $\sigma^2$.
The notation $\langle ij \rangle$ indicates a sum over nearest-neighbor
sites.  The controversy surrounded the existence of a ferromagnetic phase
transition at weak disorder in three dimensions.  Supersymmetry
techniques~\cite{ParisiGrinstein} and experiments by Hagen and
co-workers~\cite{Birgeneau} appeared to indicate that arbitrarily weak
disorder would break the system up into random domains at long enough length
scales.  Simpler arguments given by Grinstein and Ma~\cite{GrinsteinMa},
along with experiments at Santa Barbara~\cite{Jacarino} suggested that
ferromagnetism persisted until a critical value of the disorder was
reached.  The issue was settled definitively by Imbrie and
others~\cite{imbrie,bricmont}, who proved that the latter viewpoint was the
correct one.

It was realized that the unusual experimental problems posed by the
random--field model are the result of extremely slow, glassy dynamics in the
system.  Initial treatments of the dynamics~\cite{VillainGrinsteinFernandez}
concentrated on the non-equilibrium coarsening of the domains in the
ferromagnetic phase: the domain walls pin on the random fields, and as the
domains grow these pinning barriers grow too.  When the barriers become large
compared to the temperature, thermal activation becomes ineffective and the
system falls out of equilibrium without establishing long-range order.  Later
analyses by Bray and Moore~\cite{bray} and by Fisher~\cite{fisher} indicated
that this glassy behavior extends also to the equilibrium dynamics near the
critical point.  They argue that as the temperature approaches the
ferromagnetic transition temperature $t=(T-T_c(\sigma))/T_c(\sigma)\to 0$ and
the correlation length $\xi\sim t^{-\nu}$ grows, the effective coarse-grained
Hamiltonian flows to a zero temperature, disordered critical fixed point.
Unlike most critical points, the equilibrium energy scale $E\sim t^{-\theta
\nu}$ at the correlation length becomes much larger than the temperature
(violation of hyperscaling).  As a result, the temperature is an irrelevant
variable in the renormalization group (RG) treatment of the system.  The
exponent $\psi$ determining the divergence of the energy barriers $B\sim
t^{-\psi \nu}$ is assumed to equal $\theta$, which follows from the irrelevant
eigenvalue governing the RG flow of the temperature.  Instead of a
competition between bond energy and thermal fluctuations, the battle at long
length scales is between the bonds and the (renormalized) disorder.  The
dynamics, which proceeds by thermal activation with time constant $\tau \sim
\exp(B/kT)$, slows down exponentially already above $T_c(\sigma)$.

We present here a real--space RG calculation which directly confirms this
theoretical picture.  Figure~1 previews our results: the pure Ising critical
point $T_c$ is unstable to random--field disorder, and the rescaled
effective temperature and disorder flow to a disordered critical fixed point
at $T=0$, $\sigma=\sigma_c$.  As well as providing a useful qualitative
picture of the RG flows in the model, our method gives values for the
critical exponents at the two fixed points.  In addition to the simple
two-parameter truncation of Hamiltonian space, we consider additional forms
of disorder to (a)~confirm that they correspond to irrelevant operators, and
(b)~improve our critical exponents.  The calculation gives more precise
exponents than direct methods, though our systematic errors are potentially
large because of our small system size.


We investigate the model by a straightforward computational implementation
of the real--space RG~\cite{binney,niemeijer,cao} on a cubic lattice.  For
given values of the bond strength $J$, the external field $H$, and the
variance of the random fields, the procedure is as follows:
 \begin{enumerate}
 \item Choose random values for the variables $h_i$ on each site of the
lattice.
 \item Divide the lattice into cubic blocks of eight spins each.  For
each block, we define a coarse-grained spin variable.
 \item For each configuration of the coarse-grained spins, sum all the
Boltzmann factors for configurations of the original lattice consistent with
those spins, using the so-called `majority rule' (see Ref.~\cite{binney}).
 \item This defines a renormalized Hamiltonian ${\cal H}'$ which can be
inverted to give a new set of bond strengths $J'_{ij}$ and fields $h'_i$ on
the blocked lattice, as well as a number of longer-range interactions between
two or more spins.
 \item In the simplest case we discard all the longer-range interactions and
define the renormalized bond strength to be the mean $J'=\langle J'_{ij}
\rangle$ of the new bonds, the renormalized external field to be the mean
$H' = \langle h'_i \rangle$ of the new fields, and the renormalized variance
to be the variance of the new fields ${\sigma'}^2 = \langle {h'_i}^2 \rangle
- \langle h'_i \rangle^2$.  Later, we will consider more sophisticated
versions of the RG which include higher moments in the distribution of
bonds and fields than just the mean and variance considered here.  These all
turn out to be irrelevant operators, but their inclusion in the calculation
can improve the results for the critical exponents.
 \end{enumerate}
To achieve good statistics we average the values of $J'$ and ${\sigma'}^2$
over many different realizations of the randomness.  For consistency with
earlier work, we quote our results in terms of the ratios $T/J$ and $H/J$,
and the standard deviation of the random fields or `randomness' $\sigma/J$.

Employing the method first in two dimensions for a $4\times4$ system (with
rules analogous to the three-dimensional ones described above) we find no
non-trivial fixed points other than the pure Ising critical
point~\cite{nauenberg}.  We conclude that, within this approximation, there
is no phase transition at finite randomness in two-dimensions.  This is in
agreement with the findings of Imbrie~\cite{imbrie}, Berker~\cite{berker2},
Bricmont and Kupiainen~\cite{bricmont}, and others.

In three dimensions, the size of the system we can study is limited by
step~3 above, in which we are required to sum over all spin configurations
of the lattice.  We are working on more sophisticated algorithms to speed
the calculation, but for the moment we present results for a system of
$2\times2\times4$ spins.  This is large enough to give reasonable results,
but small enough for the method to be directly applicable.  Our numerical
results for the RG flows, Figure~1, are consistent with the flow diagram
postulated by Bray and Moore~\cite{bray}.  Each point in Figure~1 is an
average over $100$ different realizations of the random fields.  In the
regions close to the fixed points, we also performed a number of runs in
which we averaged over as many as $10^6$ different realizations of the
randomness in order to improve the accuracy of our values for the critical
exponents.

There are two non-trivial fixed points.  One is the normal Ising fixed point
at finite temperature and zero randomness, and the other is at zero
temperature and finite randomness.  This latter fixed point governs the
phase transition between the paramagnetic (disordered) state and the
ferromagnetic one.  As predicted by Fishman and Aharony~\cite{fishman}, all
three parameters in the problem---temperature, external field, and
randomness---are relevant at the pure Ising critical fixed point.  At the
disordered critical fixed point only the external field and the randomness
are relevant; the temperature is irrelevant.  There exists a unique
trajectory which leads from one critical fixed point to the other.  This
curve determines the crossover behavior for weakly disordered systems: far
from $T_c(\sigma)$ for small $\sigma$ the system will have critical
fluctuations given by the pure fixed point, and the growing influence of
disorder is of a universal form given by this trajectory.

The RG transformation does not actually provide us with a single
renormalized value for the bond strength $J'$, but with a distribution of
strengths, which is considerably skewed from a pure Gaussian.  We can map the
RG flows more accurately if we allow for this distribution, parameterizing
it by three quantities: the mean bond strength $J$, the standard deviation
$\sigma_J$, and the skewness $\gamma_J$.  Again, the trajectories leading
away from the fixed point are universal.

While the entire trajectory contains universal information about the critical
behavior, the most commonly measured properties are the positions of the
critical fixed points, and the critical exponents.  Our value for the
critical temperature at the pure Ising critical point is $T_c = 5.38$, which
is substantially higher than the known value of $4.51$, due to the small
system size.  Near this point, the three parameters $(T-T_c)/J$, $H/J$, and
$\sigma/J$ are expected to scale independently under the RG transformation,
and our calculation confirms this.  Our value for the critical randomness at
the disordered critical fixed point is $\sigma_c = 1.675\pm0.002$.  It has
been conjectured~\cite{nattermann} that near this fixed point the correct
eigenvectors of the RG transformation are not the bare parameters of the
problem, but are instead $T/J$, $H/J$, and the linear combination
$(\sigma-\sigma_c)/J + A(T/J)$, where A is a mixing constant.  We believe
however that the $T=0$ plane should be invariant under the RG
transformation, and this would imply that $A=0$ in general.  Our calculation
confirms this.  On the other hand, we see no physical reason why $T/J$
should be an eigenvector at the disordered critical point, and in that case
the most general scaling fields for the problem would be $H/J$,
$(\sigma-\sigma_c)/J$, and $T/J + B (\sigma-\sigma_c)/J$.  Within our
numerical calculation we find that $B=0$, but not because of any symmetry of
the Hamiltonian.  $B$ is zero because the flow is asymptotically dominated
by the lowest-lying states consistent with each configuration of
coarse-grained spins.  The argument that leads us to this result becomes
invalid in the thermodynamic limit, so it is not clear from our calculation
whether $B$ will be zero for an infinite system.

In the more sophisticated calculation, where we choose the bond strengths at
random from a skewed Gaussian distribution, we find the disordered critical
fixed point at $T = 0$, $\sigma = 1.55$, $\sigma_J = 0.36$, $\gamma_J =
-0.20$.

Table~\ref{evetable} shows our values for the eigenvalue exponents given by
linearizing the RG flows around the fixed points, together with those of
previous simulations.  Getting from the eigenvalue exponents to real critical
exponents is straightforward: for example, the barriers are thought to
diverge with exponent $\psi\nu=y_J/y_\sigma$.  We have included in the table
estimates of the statistical errors.  In addition to these statistical
errors, we expect large systematic errors because of finite-size effects on
the small lattice.  Even with these systematic errors, however, our results
at the important disordered fixed point are competitive with previous
results.  We are extending our work to a $4\times4\times4$ system, which
should allow us to include longer-range renormalized interactions and
extract reliable, accurate exponents.

The eigenvalue exponents are expected to satisfy a number of different
inequalities~\cite{fisher,berker1,schwartz,villain,young,harris}: $y_H - y_J
\le {d/2}$, $ y_H \le d$, $ y_J < {d/2}$, $ y_J \le d-1$, and $ y_H - y_J/2
> {d/2}$.  Our calculated values satisfy all of these inequalities except
for the last one, where $y_H$ is slightly less than the required value of
$\frac12(d+y_J)=2.00$.  This inequality is an expression of the fact that
the magnetization fluctuations on the phase boundary between para- and
ferromagnetic phases are expected to diverge in the limit
$q\to0$~\cite{villain}.  On the present small lattice it is not surprising
then that this inequality fails.  We hope that a calculation on a larger
lattice would bring the value of $y_H$ up.


For the renormalized system with only two coarse-grained spins, it is a
straightforward matter to calculate the distribution of barrier heights
between the four possible states of the system.  Though the small size of
the lattice probably means that the exact shape of the distribution is not
representative of the distribution in bulk systems, the scaling behavior of
the distribution with increasing length scale, which we can extract from our
RG, should be qualitatively correct.  We calculate the barrier height
distribution for a system with initial values $J$, $\sigma$ of the bond
strength and randomness, and also for systems along the RG trajectory that
starts at this point.  The successive distributions correspond to the
distribution for the original system on length scales increasing by a factor
of, in this case, two for each step along the trajectory.  In the inset to
Figure~2 we have plotted the barrier height distributions for successive
steps along one RG trajectory.  This trajectory, which is indicated by the
thicker line in Figure~1, starts fractionally below $T_c$ near the
$\sigma=0$ axis, moves along the edge of the phase boundary and lingers
briefly near the disordered critical fixed point, before turning away
towards the ferromagnetic fixed point at $T=0$, $\sigma=0$.  The barriers
are plotted in multiples of the bond strength at each stage, and the
vicinities of the fixed points are visible in the progression to longer
length scales as stationary regions in which the distribution changes little.
Figure~2 shows the mean barrier height, this time with the factors of $J$
included, as a function of length scale.  Scaling near the fixed points is
represented by the straight-line portions of this plot.  The zero slope
follows from hyperscaling; the slope 2 follows from the area of an
interface.  The structure at shorter length scales will probably not have an
effect on experiments, since it represents mean barrier heights much less
than $kT$.  But the behavior at longer scales, particularly the crossover
between the disordered and ferromagnetic fixed points might be measurable in
an experiment that correlates the size of magnetization fluctuations
(related to the length scale) with the time-scale on which they occur
(related to the barrier height).


In summary, we provide a direct implementation of the RG for the
random--field Ising model on a cubic lattice, a model whose unusual scaling
behavior was the cause of substantial controversy and which now provides the
best understood example of glassy dynamics.  Our results for the RG flows
under coarse graining and for the position and nature of the fixed points
confirm earlier conjectures.  It is encouraging that the statistical errors
on our exponents are much smaller than the errors generated by other
techniques.  If the systematic errors that arise from working on the
smallest possible lattice can be reduced by going to larger lattices, the
technique promises exponents of greater accuracy than those available at
present.  The RG also gives us a simple method for calculating the barrier
height distribution as a function of length scale.  The results may have
experimentally measurable consequences.


We would like to thank J.~F.~Marko for helpful suggestions.  This work was
partly funded by the Hertz Foundation~(BWR), the Science and Engineering
Research Council of Great Britain~(MEJN), and the NSF under grants
DMR--91--18065~(BWR,~JPS) and DMR--91--21654~(GTB).

 \begin{table}
 \begin{tabular}{l|ccc}
 Fixed point & exponent   & value           & best result     \\ \hline\hline
 Finite $T$  & $y_T$      & $1.206$         & $1.594\pm0.004$ \\
 fixed point & $y_H$      & $2.212$         & $2.488\pm0.004$ \\
             & $y_\sigma$ & $0.509\pm0.002$ & --              \\ \hline
 $T=0$       & $y_\sigma$ & $0.672\pm0.005$ & $0.9\pm0.15$    \\
 fixed point & $y_H$      & $1.88\pm0.01$   & $2.95\pm0.05$   \\
             & $y_J$      & $1.00\pm0.05$   & $1.5\pm0.2$     \\
 \end{tabular}
\caption{Results for the six eigenvalue exponents at the two critical fixed
points.  The figures in the last column are taken from Ferrenberg and
Landau~\protect\cite{ferrenberg} and Nattermann and
Villain~\protect\cite{nattermann}.}
 \label{evetable}
 \end{table}

 \begin{figure}
\caption{Lines of RG flow through Hamiltonian space for the case of zero
external field.  The range $(0, \infty)$ has been mapped onto the unit
square by taking hyperbolic tangents.  The pure Ising fixed point is
unstable to randomness:  the universal critical behavior along the
disordered phase boundary is determined by the fixed point at $(T=0,
\sigma=\sigma_c)$.  The trajectory marked with a thicker line is the one
used in the calculation of the barrier height distribution, Figure~2.}
 \label{calc}
 \end{figure}

 \begin{figure}
 \caption{The mean barrier height as a function of length scale for a system
with small randomness, slightly below the critical temperature.  The dotted
lines indicate the expected scaling of the mean barrier height in the
vicinity of the three fixed points: the slopes are zero, $1.00 \pm 0.05$,
and two, respectively.  The inset shows the actual distributions of the
barrier heights in multiples of $J$ for each point along the RG
trajectory.}
 \label{barriers}
 \end{figure}

 \end{document}